%% file: monopole-search-prd.tex
\RequirePackage{lineno} 
\documentclass[prd,twocolumn,showpacs,superscriptaddress,groupedaddress]{revtex4}  
\usepackage{graphicx}  
\usepackage{epstopdf}
\usepackage{dcolumn}   
\usepackage{bm}        
\usepackage{amssymb}   
\usepackage{subfigure}  
\usepackage{color}

\usepackage{xspace}
\def\Offline{\mbox{$\overline{\textrm%
      {Off}}$\hspace{.05em}\protect\raisebox{.4ex}%
    {$\protect\underline{\textrm{line}}$}}\xspace}%

\def\gcm{$\rm{g/cm^2}$}

\begin{document}

\leftline{Published in PRD as DOI: 10.1103/PhysRevD.94.082002}

\title{Search for ultrarelativistic magnetic monopoles with the Pierre Auger Observatory}


\input{revtex_authorlist}


\date{\today}

\begin{abstract}
We present a search for ultrarelativistic magnetic monopoles with the Pierre Auger Observatory. Such particles, possibly a relic of phase transitions in the early universe, would deposit a large amount of energy along their path through the atmosphere, comparable to that of ultrahigh-energy cosmic rays (UHECRs). The air shower profile of a magnetic monopole can be effectively distinguished by the fluorescence detector from that of standard UHECRs. No candidate was found in the data collected between 2004 and 2012, with an expected background of less than 0.1 event from UHECRs. The corresponding 90\% confidence level (C.L.) upper limits on the flux of ultrarelativistic magnetic monopoles range from  $10^{-19}$ (cm$^{2}$ sr s)$^{-1}$ for a Lorentz factor $\gamma=10^9$ to $2.5 \times10^{-21}$ (cm$^{2}$ sr s)$^{-1}$ for $\gamma=10^{12}$. These results - the first obtained with a UHECR detector - improve previously published limits by up to an order of magnitude.
\end{abstract}

\maketitle

\section{Introduction}
Maxwell's unified description of electric and magnetic phenomena is one of the greatest achievements of $19^{\rm{th}}$ century physics. Free magnetic charges and currents are not allowed in Maxwell's equations, a consequence of their apparent absence in Nature. On the other hand, there are essential theoretical motivations for magnetic monopoles. Their existence would naturally explain the quantization of electric charge, as first noted by Dirac \cite{bib:dirac} in 1931. Also, magnetic monopoles are required in Grand Unified Theories (GUTs), where they appear as intrinsically stable topological defects when a symmetry breaking results in a U(1) subgroup \cite{bib:GUTmono, bib:Giacomelli, bib:Groom}.  In typical GUT models, supermassive magnetic monopoles ($ M \approx 10^{26}$~eV/c$^2$) are produced in the early Universe at the phase transition corresponding to the spontaneous symmetry breaking of the unified fundamental interactions. When the original unified group undergoes secondary symmetry breaking at lower energy scales, so-called intermediate-mass monopoles (IMMs, $ M \sim 10^{11} - 10^{20}$~eV/c$^2$) may be generated. These particles, too massive to be produced at accelerators, may be present today as a cosmic-radiation relic of such early Universe transitions.

Supermassive magnetic monopoles should be gravitationally bound to the Galaxy 
(or to the Sun or Earth) with non-relativistic virial velocities \cite{bib:GUTmono, bib:Giacomelli, bib:Groom}.
Lighter magnetic monopoles can reach relativistic velocities through acceleration
in coherent domains of the Galactic and intergalactic magnetic fields, as well as in astrophysical objects (e.g., neutron stars)
\cite{bib:GUT-light1,bib:GUT-light2}. Kinetic energies of the order of $10^{25}$~eV have been predicted \cite{bib:wick}, which result in ultrarelativistic velocities for IMMs.
Large-exposure experimental searches for magnetic monopoles are based on their velocity-dependent interactions with matter, with a wide range of velocities allowed for GUT monopoles. 

There is a long history of experimental searches for magnetic monopoles with a variety of experiments such as MACRO \cite{bib:macro}, AMANDA \cite{bib:amanda}, Baikal \cite{bib:baikal}, SLIM \cite{bib:slim}, RICE \cite{bib:rice},  ANITA \cite{bib:anita_limit} and IceCube \cite{bib:icecube}.
The strongest upper limit on the flux of non-relativistic magnetic monopoles  ($4 \times 10^{-5} < \beta=~v/c < 0.5$) comes from the MACRO
experiment at $\approx 1.5 \times 10^{-16}$ (cm$^{2}$ sr s)$^{-1}$ (90\% C.L.) \cite{bib:macro}.  At relativistic velocities  ($\beta \approx 0.9$), the IceCube Observatory has placed the best limit at $\approx 4 \times 10^{-18}$ (cm$^{2}$ sr s)$^{-1}$ \cite{bib:icecube}. The best limit on the flux of ultrarelativistic IMMs (Lorentz factor $\gamma \approx 10^{11}$) is reported by the ANITA-II experiment 
at $\approx 10^{-19}$ (cm$^{2}$ sr s)$^{-1}$ \cite{bib:anita_limit}.

These upper limits are below the Parker bound \cite{bib:parker} of $\sim 10^{-15}$ (cm$^{2}$ sr s)$^{-1}$, which represents
 the largest possible magnetic-monopole flux consistent with survival of the Galactic magnetic field. However, the original Parker bound does not take into account the current knowledge of the Galactic magnetic field
and its almost chaotic nature, with domain lengths in the range $1 - 10$~kpc. 
The so-called ``extended Parker bound''  \cite{bib:extendedParker} becomes mass-dependent with $\Phi \sim
10^{-16} M / (10^{26} \, \rm{eV})$  (cm$^{2}$ sr s)$^{-1}$ with $M$ the monopole mass,
and is well below current experimental sensitivities (for relativistic and ultrarelativistic monopoles).

In this paper, we report a search for ultrarelativistic IMMs with data collected with the Pierre Auger Observatory between 1 December 2004 and 31 December 2012. Details of the Observatory are given in Section~\ref{sec:auger}. The search is motivated by the large energy deposited by ultrarelativistic IMMs along their path in the atmosphere, comparable to that of UHECRs, with a distinctive longitudinal development well-suited for detection by the fluorescence detector. The characteristics of air showers induced by IMMs are described in Section~\ref{sec:interaction}.  
Simulations and event reconstruction procedures are presented in Section~\ref{sec:mc}. The event selection criteria are described in Section~\ref{sec:selection}. The exposure, i.e., the time-integrated aperture, for the IMM search is evaluated in Section~\ref{sec:exposure}. Details of the data analysis and results are presented in Section~\ref{sec:analysis}. Conclusions are drawn in  Section~\ref{sec:conclusions}.

\section{Pierre Auger Observatory}
\label{sec:auger}
The Pierre Auger Observatory \cite{bib:auger} is the largest UHECR detector currently in operation. Located in the southern hemisphere in western Argentina, just northeast of the town of Malarg\"{u}e (69$^{\circ}$W, 35$^{\circ}$S, 1400 m a.s.l.), it covers an area of 3000 km$^2$ with a surface-detector array (SD)~\cite{bib:augerSD} overlooked by a fluorescence detector (FD) \cite{bib:augerFD}. 

The SD consists of 1660 water-Cherenkov detectors arranged in a triangular grid of 1500 m spacing, operating with a duty cycle of nearly 100\%. The SD stations detect at ground level the secondary particles of the extensive air shower (EAS) produced by the UHECR primary interaction in the atmosphere. 
The FD detects the UV fluorescence light from nitrogen molecules excited by the EAS particles along their path in the atmosphere.  
Its operation is limited to clear moonless nights, resulting in a duty cycle of $\sim15\%$~\cite{bib:auger}. The FD consists of 24 telescopes, arranged in groups of six at four sites overlooking the SD. Each telescope has a field of view of 30$^{\circ}$ $\times$ 30$^{\circ}$ in azimuth and elevation, with a 13 m$^2$ spherical segmented mirror collecting fluorescence light onto a 440 photomultiplier (PMT) camera.
The telescope's 3.8 m$^2$ aperture optics are of the Schmidt design and are equipped with an annular corrector lens to minimize spherical aberration.
The FD measures the longitudinal development of the UHECR shower in the atmosphere, since the fluorescence light is proportional to the energy deposited by the EAS particles~\cite{bib:fy_summary, bib:airfly1, bib:airfly2}. The depth corresponding to the maximum energy deposit, $X_{\max}$, and a calorimetric estimate of the shower energy are obtained from a fit of the shower profile.  
For the present analysis, we will use ``hybrid'' events - showers simultaneously detected by the FD and SD - which are reconstructed with superior resolution: $\sim0.6^\circ$ in arrival direction, $\sim6\%$ in energy and $\le$ 20 g/cm$^{2}$ in $X_{\max}$, respectively~\cite{bib:auger_composition}. 
Systematic uncertainties on the energy and $X_{\max}$ are 14\%~\cite{bib:auger, bib:EnergyScaleICRC13} and $\le$10 g/cm$^2$~\cite{bib:auger_composition}, respectively. 

\section{Ultrarelativistic Monopole-induced Air showers}
\label{sec:interaction}
\begin{figure}
  \begin{center}
    \includegraphics[width=1.0\linewidth]{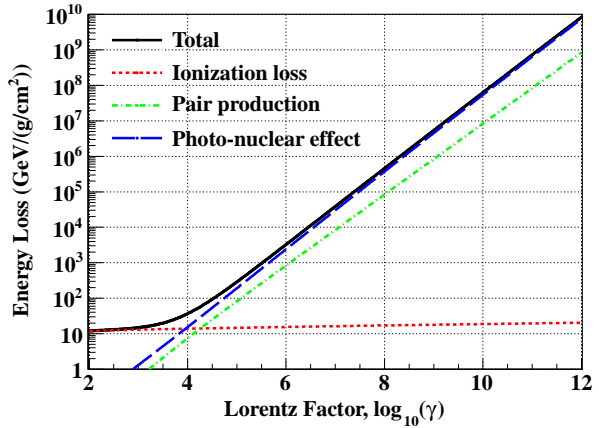}
    \caption{Energy loss of a magnetic monopole in air as a function of its Lorentz factor $\gamma$. }
    \label{fig:eloss}
  \end{center}
\end{figure}

Electromagnetic interactions of magnetic monopoles have been extensively investigated~\cite{bib:wick,bib:eminteract}.  The electromagnetic energy loss of a magnetic monopole in air is shown in Figure~\ref{fig:eloss} as a function of its Lorentz factor $\gamma = E_{\rm{mon}}/M$.  Collisional energy loss is the dominant contribution for $\gamma \le 10^4$. At higher Lorentz factors, pair production and photo-nuclear interactions become the main cause of energy loss. Bremsstrahlung is highly suppressed by the large monopole mass. An ultrarelativistic IMM would deposit a large amount of energy in its passage through the Earth's atmosphere, comparable to that of a UHECR.  For example, a singly-charged IMM with $\gamma = 10^{11}$ loses  $\approx$ 700~PeV/(g/cm$^2$) (cf. Figure~\ref{fig:eloss}), which sums up to  $\approx 10^{20.8}$~eV when integrated over an atmospheric depth of $\approx 1000$~g/cm$^2$. This energy will be dissipated by the IMM through production of secondary showers initiated by photo-nuclear effects and pair productions along its path. 

In order to study the characteristics of IMM-induced showers, we implemented magnetic-monopole interactions in the CORSIKA air-shower simulation software \cite{bib:corsika}. 
Specifically, existing subroutines for muonic collisional loss, $e^+e^-$-pair production and photo-nuclear interaction were appropriately modified in CONEX \cite{bib:Conex}, which can be used within CORSIKA to perform a combination of stochastic particle production and numeric integration of particle cascades.
We used \cite{bib:pairprod,bib:pairprod1} to parameterize the differential cross section for $e^+e^-$-pair production and the Bezrukov-Bugaev parameterization \cite{bib:photonucl,bib:photonucl1} for the photo-nuclear interaction model. 
To describe magnetic monopole interactions, the cross sections were scaled up by a factor $z_M^2$ \cite{bib:wick,bib:eminteract}, where $z_M=1/(2\alpha)$ is the singly-charged monopole charge and $\alpha$ is the fine-structure  constant. Pair production and photo-nuclear interactions were treated explicitly as stochastic processes resulting in secondary particles produced along the monopole path in the atmosphere. Standard CONEX routines were used to simulate showers originating from these secondary particles. Collisional losses were implemented as continuous energy losses.

\begin{figure}
  \begin{center}
    \includegraphics[width=1.0\linewidth]{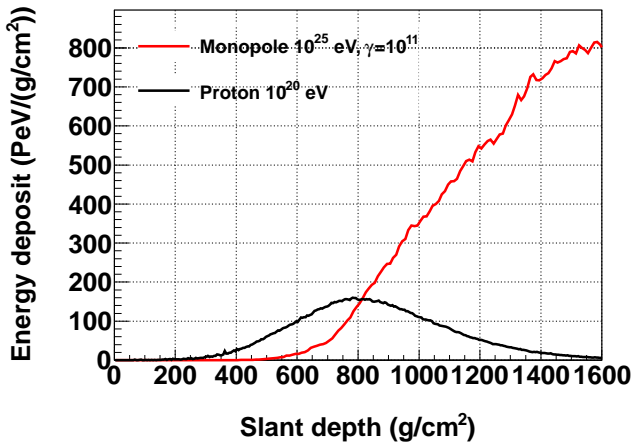}
    \caption{Longitudinal profile of the energy deposited by an ultrarelativistic IMM of  $E_{\rm{mon}}= 10^{25}$~eV, $\gamma = 10^{11}$ and zenith angle of $70^\circ$ (red solid line). The profile of a UHECR proton shower of energy $10^{20}$~eV is shown as a black solid line.}
    \label{fig:monopoleShower}
  \end{center}
\end{figure}

The longitudinal profile of the energy deposited by an ultrarelativistic IMM of  $E_{\rm{mon}}= 10^{25}$~eV, $\gamma = 10^{11}$ and zenith angle of $70^\circ$ is shown in  Figure~\ref{fig:monopoleShower}. When compared with a standard UHECR proton shower of energy $10^{20}$~eV (black solid line in Figure~\ref{fig:monopoleShower}), the IMM shower presents a much larger energy deposit and deeper development, due to the superposition of many showers uniformly produced by the IMM along its path in the atmosphere. This distinctive feature will be used in our analysis, which is based on the shower development measured in the hybrid events.
Also, we have confirmed this feature in case if we use other parameterizations (e.g., ALLM \cite{bib:allm}), meaning the difference between cross sections is a second order effect for the shower profile of IMM.
Depending on their energy, ultrarelativistic IMMs may traverse the Earth \cite{bib:anita_limit,bib:icecube} and emerge from the ground producing upward-going showers. We have not searched for this kind of candidate, which would not guarantee a high-quality reconstruction of the shower development.

\section{Monte Carlo Simulations and Event Reconstruction}
\label{sec:mc}
Monte Carlo samples of ultrarelativistic IMMs were simulated for Lorentz factors in the range $\gamma=10^8 -10^{12}$ at a fixed monopole energy of $E_{\rm{mon}} = 10^{25}$~eV,  
because the monopole energy loss does not depend on $E_{\rm{mon}}$ but rather on $\gamma$ in the ultrarelativistic regime of this search. 
While we used a fixed $E_{\rm{mon}}$ in the simulations, the results can be readily applied to a much larger range of monopole energies.

To estimate the background from UHECRs, we simulated proton showers with energy $E_{\rm{p}}$ between $10^{18}$~eV and $10^{21}$~eV. 
Proton primaries are chosen to obtain a conservative estimate of the cosmic-ray background (cf. Sec.~\ref{sec:analysis}).
We used three different models - QGSJetII-04, Sibyll 2.1 and EPOS-LHC - to account for uncertainties in the hadronic interactions. Events were simulated according to an $E_{\rm{p}}^{-1}$  energy spectrum, to ensure sufficient Monte Carlo statistics at the highest energy, and then appropriately weighted to reproduce the energy spectrum measured by the Pierre Auger Observatory \cite{bib:AugerICRC13}.  

\begin{figure}
  \begin{center}
    \includegraphics[width=0.95\linewidth]{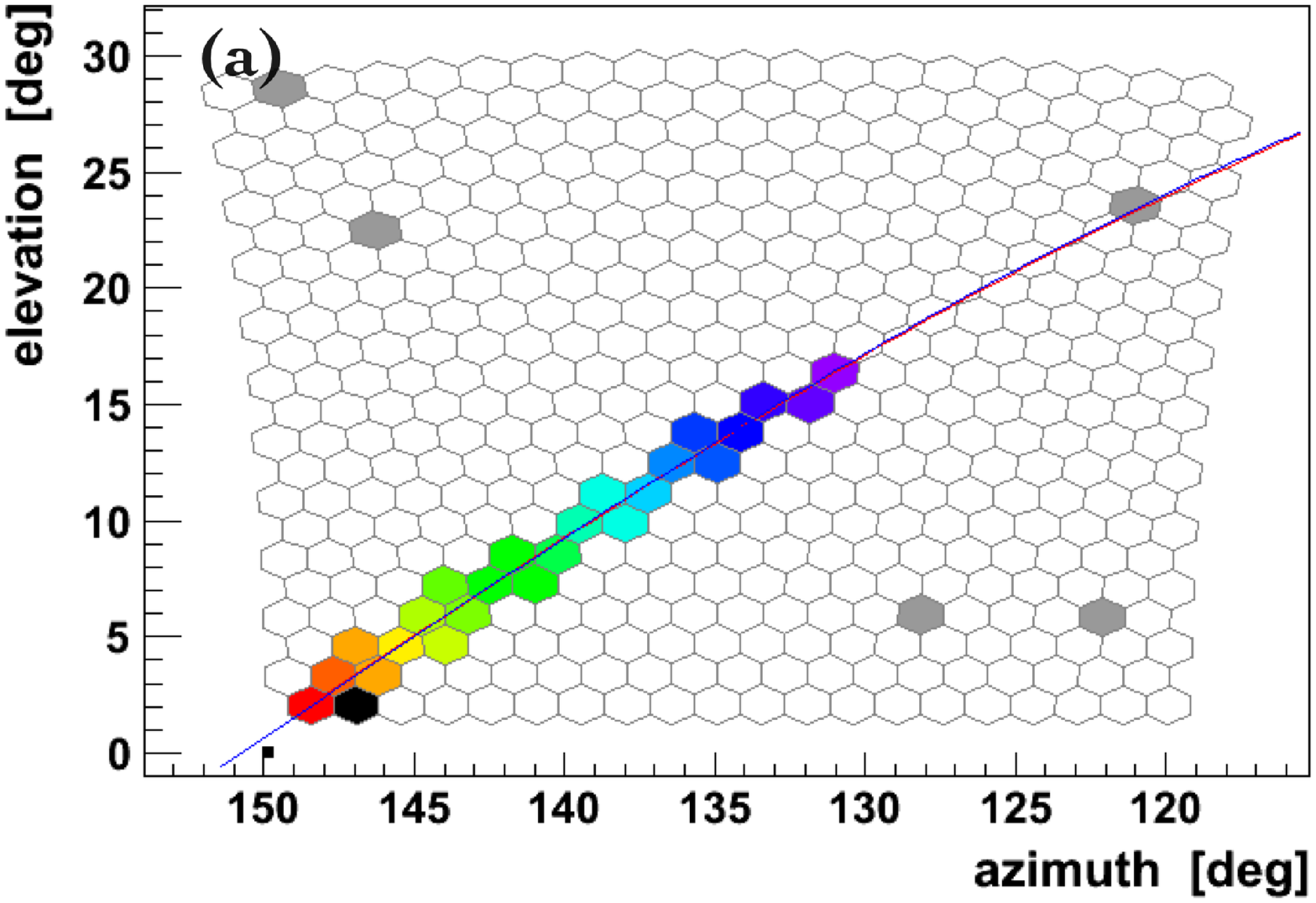}
     \includegraphics[width=1.0\linewidth]{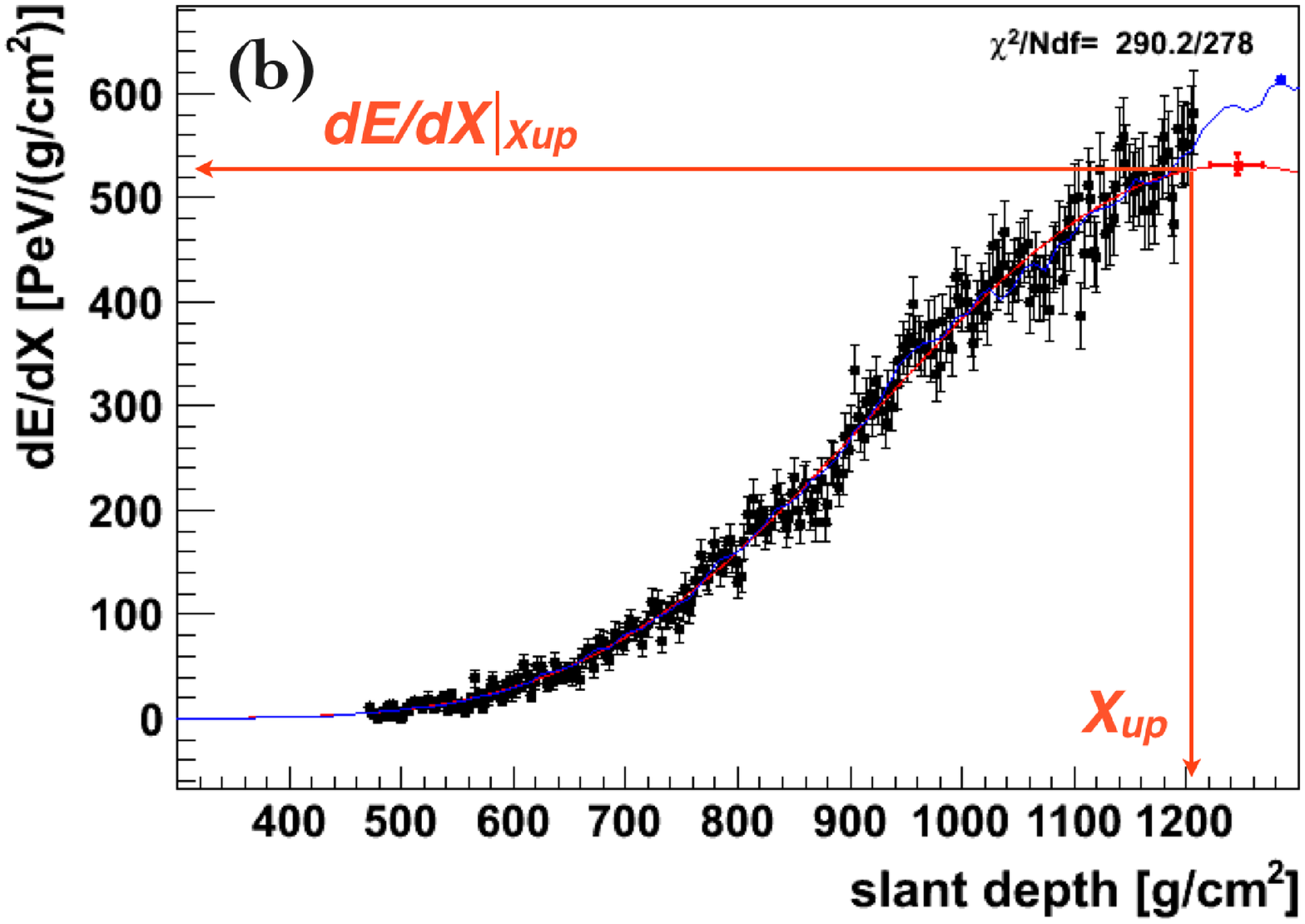}
    \caption{Reconstructed signals for a simulated magnetic monopole of energy $10^{25}$~eV  and $ \gamma=10^{11}$. In (a), the FD camera view is shown with color-coded timing of triggered pixels (time increases from blue to red). The red (blue) line indicates the reconstructed (simulated) shower direction projected on the camera view.  In (b), the reconstructed longitudinal profile of the shower is shown. The red line is the result of a Gaisser-Hillas fit of the profile, with the red cross indicating the position of $X_{\max}$. The blue line represents the simulated profile of the monopole shower. The selection variables $X_{\rm{up}}$, the largest visible slant depth, and $dE/dX|_{X\rm{up}}$, energy deposited at $X_{\rm{up}}$, are also indicated. } 
    \label{fig:monopoleSimulation}
  \end{center}
\end{figure}

For both the IMM and UHECR simulations, we used the CORSIKA package \cite{bib:corsika} to generate an isotropic distribution of showers above the horizon, and the Auger \Offline software \cite{bib:auger_fdanalysis} to produce the corresponding FD and SD events.
We found that the standard event reconstruction, which is optimized for UHECRs, provides equally accurate direction and longitudinal profile for ultrarelativistic IMM showers.
An example of reconstructed longitudinal profile for a simulated magnetic monopole of energy $10^{25}$~eV  and $\gamma=10^{11}$ is shown in Figure~\ref{fig:monopoleSimulation} indicating the profile of the generated CORSIKA shower (blue line) and the result of a fit of the reconstructed profile with a Gaisser-Hillas function~\cite{bib:gh_function} (red line). For standard UHECRs, the energy, $E_{\rm{sh}}$, and the depth of maximum development, $X_{\rm{max}}$, of the shower are estimated by the integral of the fitted profile and by the position of its maximum, respectively. 
When applied to an ultrarelativistic IMM shower profile, the Gaisser-Hillas parameterization provides a very good fit of the portion of the profile detected in the FD field of view (cf. red and blue lines in Figure~\ref{fig:monopoleSimulation} in the relevant range). 
Also, due to the steep rising of the ultrarelativistic IMM profile, the fit systematically converges to a value of $X_{\rm{max}}$ beyond the lower edge of the FD field of view, corresponding to the largest visible slant depth, $X_{\rm{up}}$. 
We will use this characteristic to reject most of the standard UHECR showers, which constitute the background for this search. Since $X_{\rm{max}}$ of standard UHECR showers are located in FD field of view, a specific selection is required to search for the IMM profile.

\section{Event Selection}
\label{sec:selection}
We restricted our event selection to time periods with good operating conditions of the FD telescopes and well-defined calibration constants. Additional requirements were imposed on the quality of the atmosphere (aerosols and cloud coverage). Details on these data-quality criteria can be found in~\cite{bib:auger_composition}.  A total of 376,084 hybrid shower candidates were selected.

A further set of selection criteria was applied to ensure good-quality showers. We required the zenith angle of the shower to be $<60^\circ$, and the distance of the shower core to the SD station with the highest signal to be less than 1500 m. The shower must be seen by at least five FD pixels over a slant depth interval of at least 200~\gcm. We rejected events with gaps in their profile of more than 20\% of the profile length, which could be due to telescope-border effects. 
The Gaisser-Hillas fit of the shower profile was required to have a $\chi^2 / \rm{ndf}<2.5$, where ndf is the number of degrees of freedom. To guarantee full SD-trigger efficiency, the shower must have a minimum energy. Rather than using $E_{\rm{sh}}$, which is ill-defined for an ultrarelativistic IMM shower, we employed the energy deposited at the largest visible slant depth $X_{\rm{up}}$, $dE/dX|_{X\rm{up}}$, as a discriminating variable related to the shower energy (Figure~\ref{fig:monopoleSimulation}). The $dE/dX|_{X\rm{up}}$ is calculated by the result of the Gaisser-Hillas fit. The requirement $dE/dX|_{X\rm{up}}>$ 3.0~PeV/(g/cm$^2$) is equivalent to an energy threshold of $\approx 10^{18.5}$~eV, where the SD is fully efficient. These shower-quality criteria selected a sample of well-reconstructed events, and are efficient for UHECRs as well as ultrarelativistic IMM showers.

\begin{figure}
  \begin{center}
    \subfigure[Shower quality selection]{\includegraphics[width=1.0\linewidth]{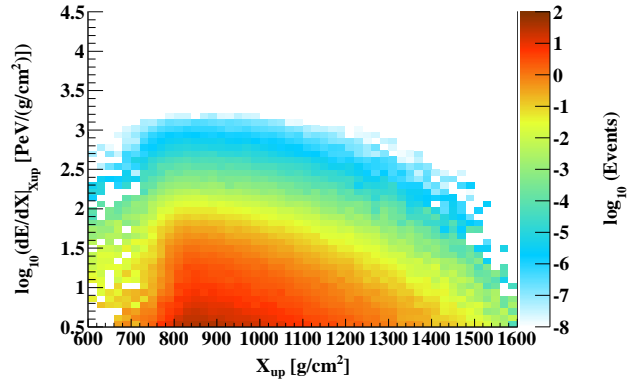}}
    \subfigure[$X_{\rm{max}}>X_{\rm{up}}$ selection]{\includegraphics[width=1.0\linewidth]{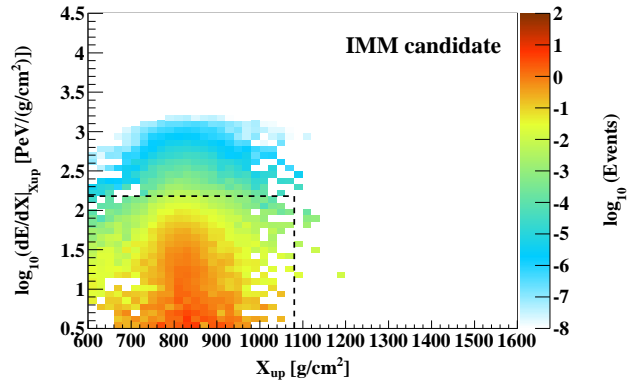}}
    \caption{Correlation of $dE/dX|_{X\rm{up}}$ with $X_{\rm{up}}$ for simulated UHECR proton showers passing the quality-selection criteria (a) and the additional requirement $X_{\rm{max}}>X_{\rm{up}}$ (b). The color-coded scale indicates the number of events expected in the search-period data set based on the energy spectrum measured with Auger~\cite{bib:AugerICRC13}. Only events outside the dashed box in (b) are kept in the final selection for ultrarelativistic IMMs. 
}
    \label{fig:ProtonBackground}
  \end{center}
\end{figure}

\begin{figure}
  \begin{center}
    \subfigure[$\gamma=10^{10}$]{\includegraphics[width=1.0\linewidth]{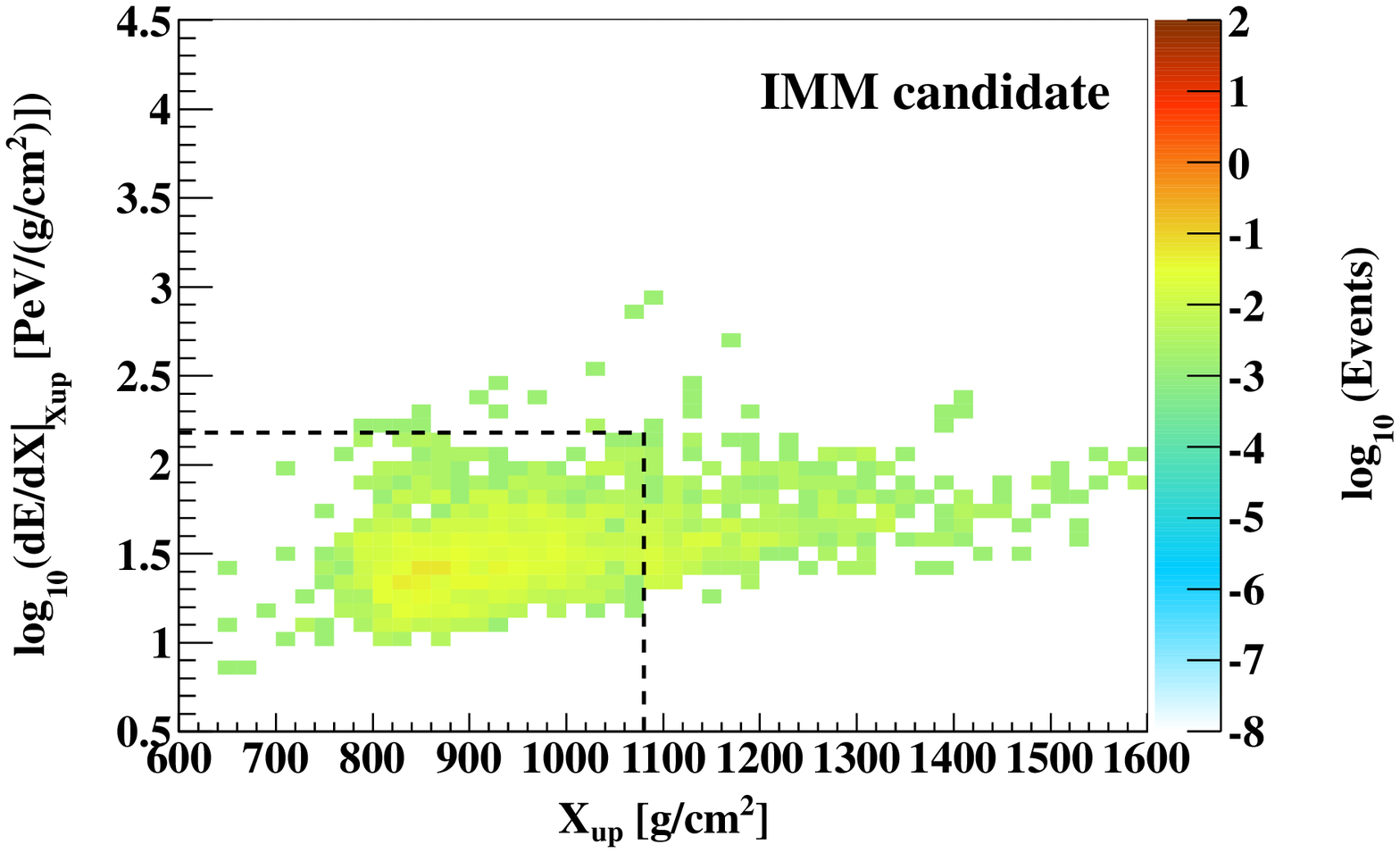}}
    \subfigure[$\gamma=10^{11}$]{\includegraphics[width=1.0\linewidth]{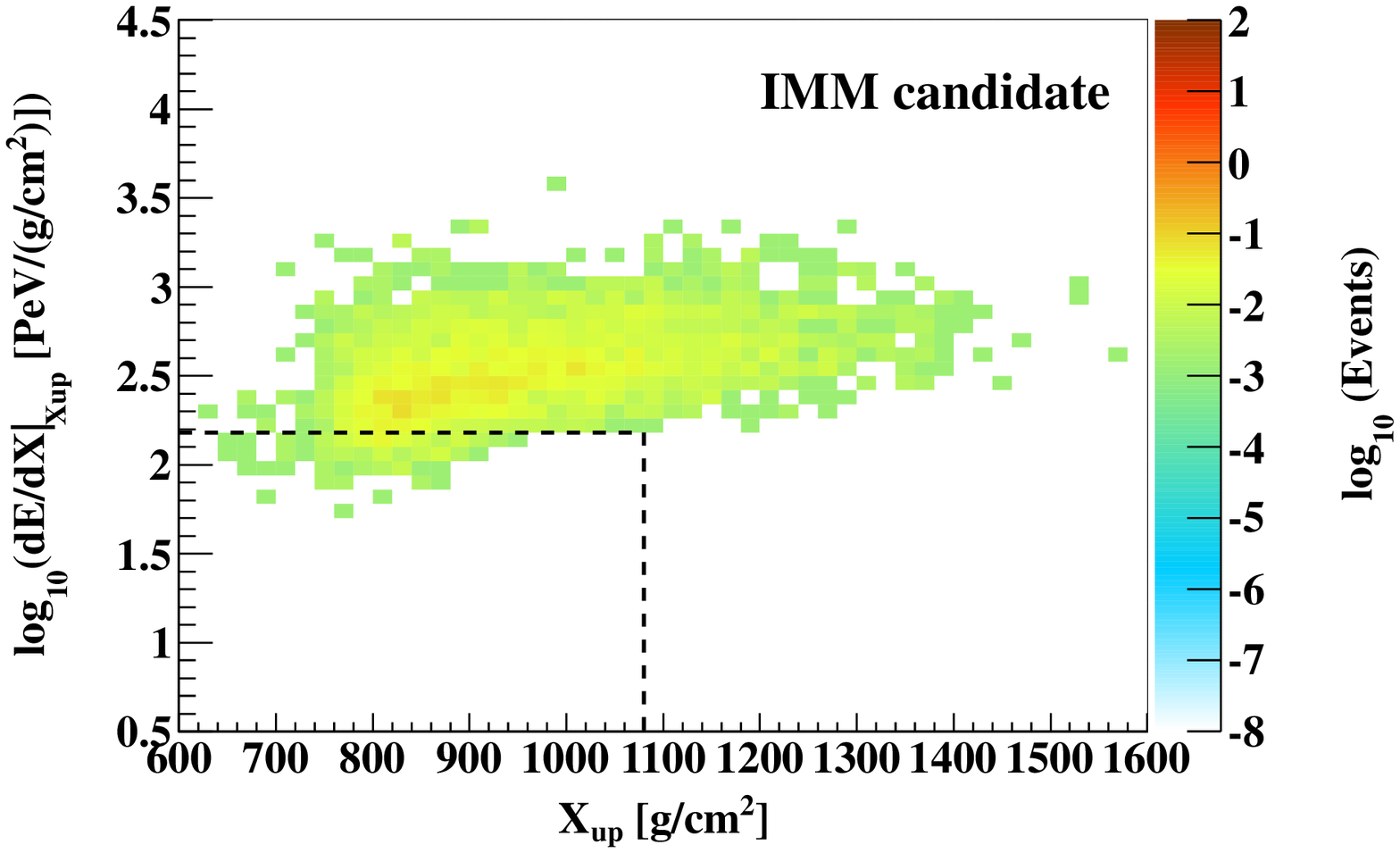}}
    \caption{Correlation of $dE/dX|_{X\rm{up}}$ with $X_{\rm{up}}$ for simulated ultrarelativistic IMM of energy $10^{25}$~eV and Lorentz factors $\gamma=10^{10}$(a) and $10^{11}$(b). The color-coded scale indicates the number of events expected in the search-period data set assuming a flux of $10^{-20}$~(cm$^{2}$ sr s)$^{-1}$. Only events outside the dashed boxes are kept in the final selection for ultrarelativistic IMMs.} 
    \label{fig:MonopoleSignal}
  \end{center}
\end{figure}

Additional criteria for IMM selection were established from Monte Carlo simulations described in Section~\ref{sec:mc}.  We required $X_{\rm{max}}$ to be  larger than $X_{\rm{up}}$, which is almost always fulfilled by ultrarelativistic IMM showers. Only  6\% of the UHECR proton showers of $10^{18.5}$~eV survived this cut, the fraction increasing to 32\% for $10^{20.5}$~eV showers. A further reduction was obtained by appropriate constraints on the penetration of the shower and its energy deposit. To illustrate this second requirement, we show in Figure~\ref{fig:ProtonBackground}(a) the correlation of $dE/dX|_{X\rm{up}}$ with $X_{\rm{up}}$ for UHECR background events passing the shower-quality criteria. When $X_{\rm{max}}>X_{\rm{up}}$ is required, the number of events is drastically reduced and the population becomes constrained in a much smaller region, as shown in Figure~\ref{fig:ProtonBackground}(b). The maximum value of $X_{\rm{max}}$  found in the UHECR proton simulated events is $\approx 1100$~g/cm$^2$, which results in the $X_{\rm{up}}$ upper boundary of Figure~\ref{fig:ProtonBackground}(b): $X_{\rm{max}}$ is always in the FD field of view when $X_{\rm{up}} \gtrsim 1100$~g/cm$^2$. 
On the other hand, the reconstructed $X_{\rm{max}}$ will always be outside the FD field of view for ultrarelativistic IMM showers, independently of the shower's $X_{\rm{up}}$. This is apparent in Figure~\ref{fig:MonopoleSignal}, where  the correlation of $dE/dX|_{X\rm{up}}$ with $X_{\rm{up}}$ is shown for ultrarelativistic IMM simulated events. 
The background from UHECRs is almost eliminated by excluding an appropriate region of the ($X_{\rm{up}}$, $dE/dX|_{X\rm{up}}$) plane. We optimized the selection to achieve less than 0.1 background event expected in the data set of this search. The final requirement, $X_{\rm{up}}>1080$~g/cm$^2$ or $dE/dX|_{X\rm{up}}>150$~ PeV/(g/cm$^2$), is shown in Figure~\ref{fig:ProtonBackground}(b) and Figure~\ref{fig:MonopoleSignal} as dashed boxes, and results in an expected background of 0.07 event in the search-period data set.

The selection criteria used for this search are summarized in Table~\ref{tbl:eventSelection}. The corresponding selection efficiency for ultrarelativistic IMMs ranges from 3\% for  $\gamma = 10^9$ to 91\% for $\gamma = 10^{12}$ (see Table~\ref{tbl:uplimit}).

\begin{table}
 \begin{tabular}{lrr}
    \hline
    Shower-quality selection criteria      & \#events  & $f$(\%) \\ 
    \hline
    Reconstructed events                   &  376,084  &   ---  \\
    Zenith angle  $<60^\circ$              &  360,159  &  95.8 \\    
    Distance from nearest SD $<$ 1500 m    &  359,467  &  99.8 \\
    Number of FD pixels $>$ 5              &  321,293  &  89.4 \\
    Slant-depth interval  $>$ 200 g/cm$^2$ &  205,165  &  63.9 \\
    Gaps in profile $<$ 20\%               &  199,625  &  97.3 \\
    profile fit $\chi^2$/ndf $<$ 2.5       &  197,293  &  98.8 \\   
    $dE/dX|_{X\rm{up}}$       $>$ 3.0 PeV/(g/cm$^2$) & 6812 & 3.5  \\
    \hline
    Magnetic-monopole selection criteria  \\ 
    \hline  
    $X_{\max} > X_{\rm{up}}$ & 352 & 5.2 \\
    $X_{\rm{up}}>1080$~g/cm$^2$ or & & \\
    $dE/dX|_{X\rm{up}}>150$~ PeV/(g/cm$^2$) & 0 & 0.0 \\
     \hline
 \end{tabular}
\caption{Event-selection criteria and data-selection results. The number of events passing each selection criterion is reported, together with the corresponding fraction of events remaining, $f$.}
 \label{tbl:eventSelection}
\end{table}

\section{Exposure}
\label{sec:exposure}
The flux $\Phi$ of ultrarelativistic IMMs of Lorentz factor $\gamma$ is given by 
\begin{equation}
\Phi(\gamma)=\frac{k}{\mathcal{E}(\gamma)},
\label{eq:upperlimit}
\end{equation}
where $k$ is the number of events surviving the selection criteria of Table~\ref{tbl:eventSelection} (or an appropriate upper limit if no candidate is found), and  $\mathcal{E}(\gamma)$ is the exposure, i.e., the time-integrated aperture for the hybrid detection of ultrarelativistic IMMs. The exposure is defined as \cite{bib:hybrid_expo}:
\begin{equation}
\mathcal{E}(\gamma) =  \int_{S_{\rm{gen}}} \int_{\Omega} \int_T \epsilon (\gamma,t,\theta,\phi,x,y) \cos \theta dS d\Omega dt,
\label{eq:expo}
\end{equation}
where $\epsilon$ is the detection efficiency for an ultrarelativistic IMM of zenith angle $\theta$ and azimuth angle $\phi$ intersecting the ground at a position $(x,y)$, $\Omega$ is the solid angle, $S_{\rm{gen}}$ is the area over which events are detectable, and $T$ is the time period of the search data set.  

In general, the detection efficiency $\epsilon$ changes over time, which must be taken into account in the calculation of the exposure. In fact, the effective area of the SD array and the number of operating FD telescopes grew during the Observatory installation from 2004 to 2008, and then varied due to occasional failures of the SD stations or FD telescopes. Sometimes weather conditions (e.g., wind, rain) introduced down-time in the operation of the FD.  Also, the night-sky background and atmospheric conditions, such as aerosol concentration and cloud coverage, changed during data taking, which affected the sensitivity of the FD telescopes.

These effects were properly taken into account with a time-dependent detector simulation \cite{bib:hybrid_expo}, which makes use of slow-control information and atmospheric measurements recorded during data taking. The detector configuration and atmospheric characteristics were changed in the simulation according to the time period $T$.  For each Lorentz factor $\gamma$, we generated a number  $N(\gamma,\cos \theta)$ of ultrarelativistic IMM showers over an area $S_\mathrm{gen}$, with $n(\gamma, \cos \theta)$ of them fulfilling the event-selection criteria of Table~\ref{tbl:eventSelection}. Then the exposure given by Equation~\ref{eq:expo} was numerically evaluated:
\begin{equation}
\mathcal{E}(\gamma)  = 
  2 \pi ~ S_\mathrm{gen} ~ T ~ 
   \sum_{i} \frac{n(\gamma, \cos \theta_{i})}{N(\gamma,\cos \theta_{i})} ~ \cos \theta_{i} ~ \Delta \cos \theta_{i}.
\label{eq:discr_exposure}
\end{equation}
Table~\ref{tbl:uplimit} shows the estimated hybrid exposure as a function of the IMM Lorentz factor. The exposure corresponding to the search period ranges from $\approx 100~\rm{km^{2}}$ sr yr for $\gamma=10^9$ to  $\approx 3000~\rm{km^{2}}$ sr yr for $\gamma \ge 10^{11}$. 
Several sources of systematic uncertainties were considered. The uncertainty of the on-time calculation resulted in an uncertainty of 4\% on the exposure. The detection efficiency estimated through the time-dependent detector simulation depends on the fluorescence yield assumed in the simulation, on the FD shower-reconstruction methods and on the atmospheric parameters and FD calibration constants recorded during data taking. Following the procedures of \cite{bib:hybrid_expo}, the corresponding uncertainty on the exposure was estimated to be 18\%. To estimate the uncertainty associated with the event selection, we changed the size of the ($X_{\rm{up}}$, $dE/dX|_{X\rm{up}}$) selection box according to the uncertainty on the two selection variables. $X_{\rm{up}}$ was changed by $\pm10$~g/cm$^2$, corresponding to the uncertainty on $X_{\max}$~\cite{bib:auger_composition}, and $dE/dX|_{X\rm{up}}$ was changed by the uncertainty on the FD energy scale~\cite{bib:AugerICRC13}. The number of selected IMM events changed by 9\%, which was taken as an estimate of the uncertainty on the exposure. From the sum in quadrature of these uncertainties, a total systematic uncertainty of 21\% was assigned to the exposure. 

\section{Data Analysis and Results}
\label{sec:analysis}
\begin{figure}
  \begin{center}
    \includegraphics[width=1.0\linewidth]{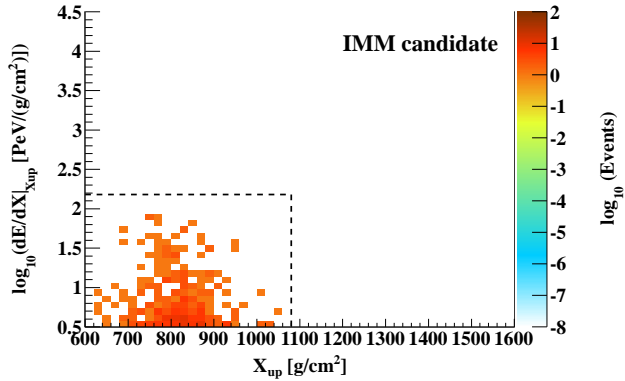}
    \caption{Correlation of $dE/dX|_{X\rm{up}}$ with $X_{\rm{up}}$ for the data sample passing the shower-quality selection criteria and $X_{\max} > X_{\rm{up}}$. The color-coded scale indicates the number of events.
 No event is found outside the dashed box in the final selection for ultrarelativistic IMMs.}
    \label{fig:DataResult}
  \end{center}
\end{figure}

\begin{figure}
  \begin{center}
    \includegraphics[width=1.0\linewidth]{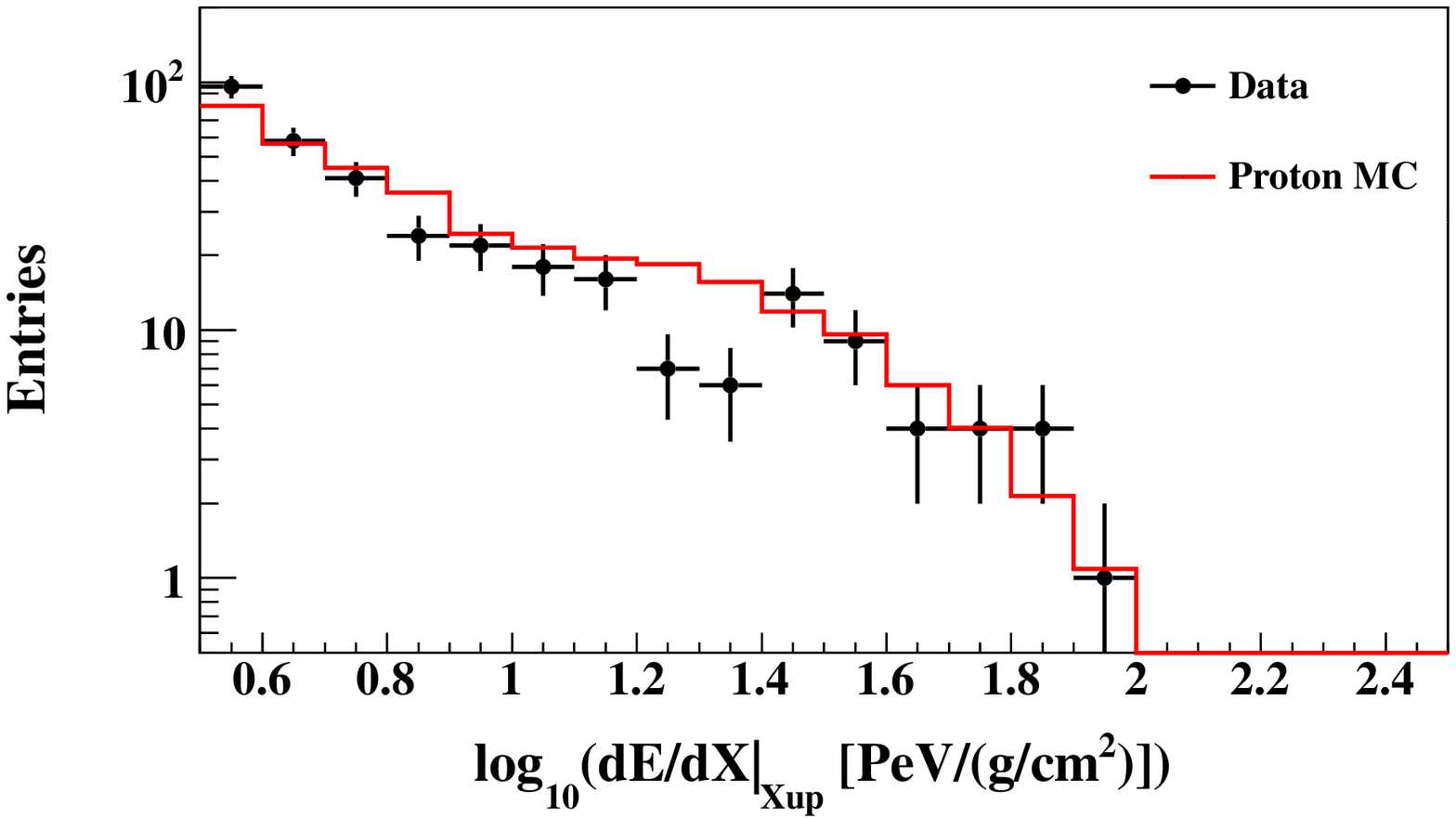}
    \includegraphics[width=1.0\linewidth]{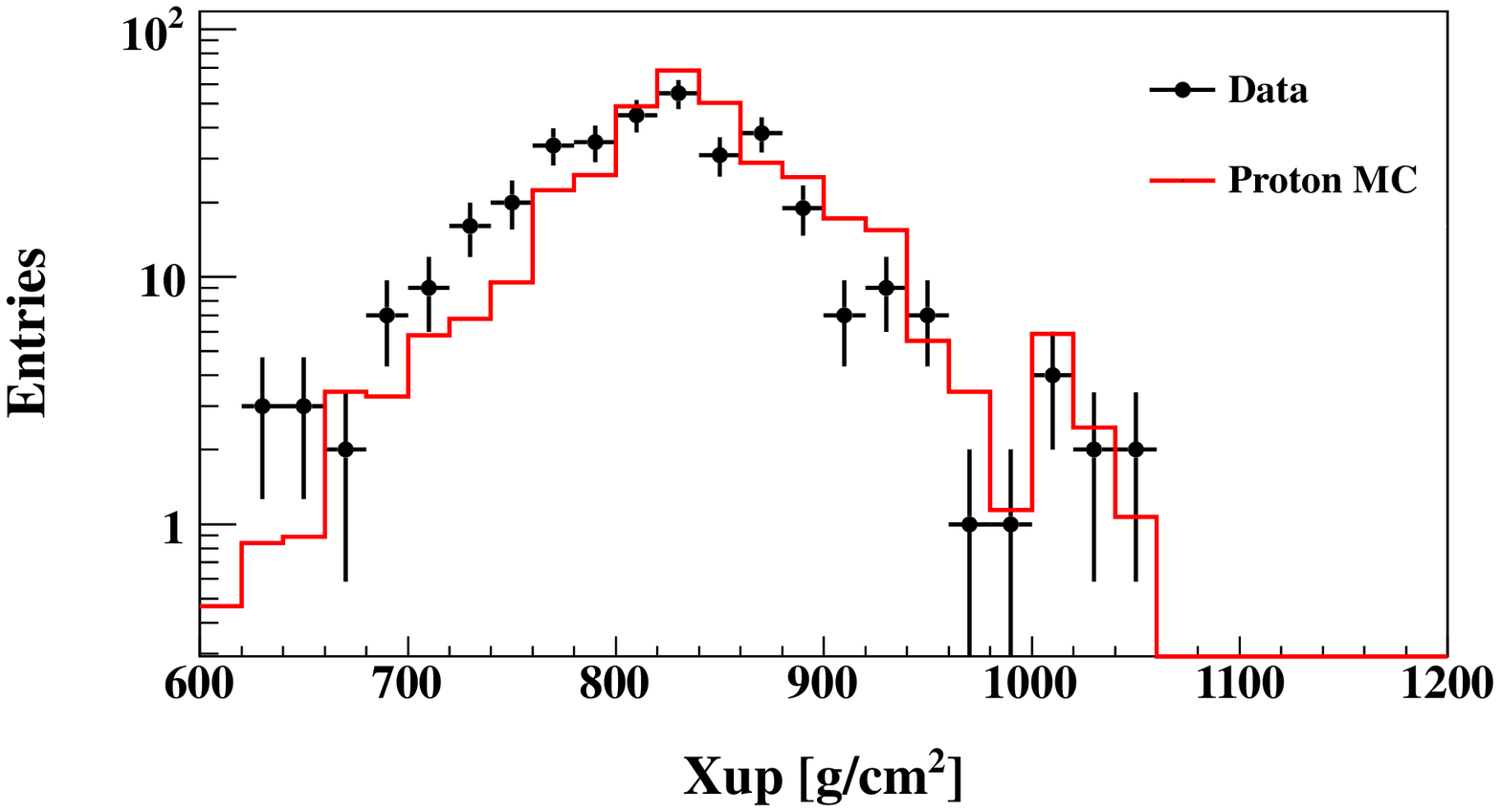}	
    \caption{Distribution of $dE/dX|_{X\rm{up}}$ (a) and $X_{\rm{up}}$ (b) for the data sample (black dots) passing the shower-quality selection criteria and $X_{\max} > X_{\rm{up}}$. The red solid line is the Monte Carlo prediction for a pure UHECR proton background, normalized to the number of selected events in the data.}
    \label{fig:distrib}
  \end{center}
\end{figure}

The search for ultrarelativistic IMMs was performed following a blind procedure. The selection criteria described in Section~\ref{sec:selection} were optimized using Monte Carlo simulations and a small fraction (10\%) of the data. This training data set was excluded from the final search period. Then the selection was applied to the full sample of data collected between 1 December 2004 and 31 December 2012. The number of events passing each of the selection criteria is reported in Table~\ref{tbl:eventSelection}. The correlation of $dE/dX|_{X\rm{up}}$ with $X_{\rm{up}}$ for events passing the shower-quality criteria and $X_{\max} > X_{\rm{up}}$  is shown in Figure~\ref{fig:DataResult}. The corresponding distributions of $dE/dX|_{X\rm{up}}$ and $X_{\rm{up}}$ are compared in  Figure~\ref{fig:distrib} with Monte Carlo expectations for a pure UHECR proton background, showing a reasonable agreement between data and simulations. The partial difference indicates there are heavier nuclei than protons as well.
 No event passed the final requirement in the ($X_{\rm{up}}$, $dE/dX|_{X\rm{up}}$) plane, and the search ended with no candidate for ultrarelativistic IMMs. 

Given the null result of the search, a 90\% C.L. upper limit on the flux of ultrarelativistic IMMs, $\Phi_{90\%\rm{C.L.}}$, was derived from Equation~\ref{eq:upperlimit}, with exposure $\mathcal{E}(\gamma)$ as in Table \ref{tbl:uplimit} and $k=2.44$. This value of $k$ corresponds to the Feldman-Cousins upper limit \cite{bib:feldman} for zero candidates and zero background events.  We derived in Section~\ref{sec:selection} a background level of 0.07 event which is likely to be overestimated, since a pure proton composition was assumed while heavier nuclei appear to be a dominant component at the highest energies~\cite{bib:auger_composition}. In fact, the fraction of deeply-penetrating showers produced by heavy nuclei is significantly smaller resulting in fewer background events for the IMM search.  Given the uncertainty in the background, we have taken a conservative approach and assumed zero background events, which provides a slightly worse limit.  

In Section~\ref{sec:exposure} we estimated a 21\% systematic uncertainty on the exposure which must be taken into account in the upper limit. 
Rather than following the propagation of statistical and systematic uncertainties outlined in~\cite{bib:Cousins_Highland}, which would worsen the upper limit by a factor of 1.05, we adopted a more conservative approach and multiplied $\Phi_{90\%\rm{C.L.}}$ by a factor of $f = 1 + n \times 0.21$, where $n = 1.28$ corresponds to the 90\% C.L. 

\begin{figure}
  \begin{center}
    \includegraphics[width=1.0\linewidth]{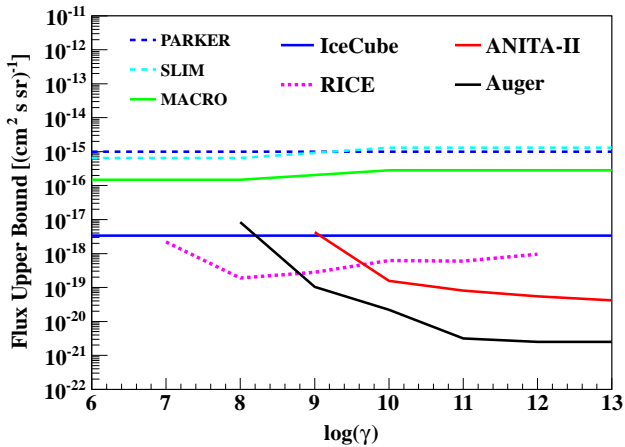}
    \caption{90\% C.L. upper limits on the flux of ultrarelativistic IMMs:  this work (black solid line); Parker bound (blue dashed line) \cite{bib:parker}; SLIM (sky-blue dashed line) \cite{bib:slim}, MACRO (green solid line) \cite{bib:macro}, IceCube (blue solid line) \cite{bib:icecube}, RICE (pink dotted line) \cite{bib:rice} and ANITA-II (red line) \cite{bib:anita_limit}. The MACRO and SLIM limits above $\gamma =10^9$ were weakened by a factor of two to account for the IMM attenuation through the Earth.}
    \label{fig:limit}
  \end{center}
\end{figure}

Our final 90\% C.L. upper limits on the flux of ultrarelativistic IMMs are reported in Table~\ref{tbl:uplimit} and shown in Figure~\ref{fig:limit}, together with results from previous experiments. Following the treatment of \cite{bib:anita_limit}, the MACRO and SLIM limits extrapolated to $\gamma \ge 10^9$ were weakened by a factor of two to account for the IMM attenuation when passing through the Earth. 

\begin{table}
\begin{tabular}{c|c|c}
     $\log_{10}(\gamma)$ & $\mathcal{E}(\gamma)$ (km$^2$ sr yr) & $\Phi_{90\%\rm{C.L.}}$ (cm$^2$ sr s)$^{-1}$ \\ 
     \hline
     8         & 1.16                  & 8.43 $\times 10^{-18}$ \\
     9         & 9.52 $\times 10^{1}$  & 1.03 $\times 10^{-19}$ \\
     10        & 4.50 $\times 10^{2}$  & 2.18 $\times 10^{-20}$ \\
     11        & 3.15 $\times 10^{3}$  & 3.12 $\times 10^{-21}$ \\
     $\ge$ 12  & 3.91 $\times 10^{3}$  & 2.51 $\times 10^{-21}$ \\
 \end{tabular}
 \caption{ Exposure and 90\% C.L. upper limits on the flux of ultrarelativistic IMMs ($E_{\rm{mon}}=10^{25}$~eV) for different Lorentz factors $\gamma$. A 21\% Systematic uncertainty on the exposure was taken into account in the upper limits.}
 \label{tbl:uplimit}
\end{table}

Several checks of the analysis were performed. 
Variation of the selection criteria within reasonable ranges still resulted in no candidate. 
The UHECR energy spectrum was varied within its uncertainties \cite{bib:AugerICRC13}, with negligible effect on the background estimation. The background for the IMM search is dominated by deeply-penetrating UHECR showers, which are found in the tail of the $X_{\max}$ distribution and depend on the characteristics of the hadronic interactions. We used three different hadronic-interaction models (Section~\ref{sec:selection}) to simulate UHECR protons for background estimation. 
Ultrahigh-energy photons are also expected to produce deeply-penetrating showers, which may mimic an IMM event. The photon hypothesis should be carefully evaluated in case a candidate IMM is found. Since this search ended with a null result, the zero background assumption produces the most conservative limit also including the possibility of ultrahigh-energy photons.
Lastly, we compared the CORSIKA energy-loss model with analytical approximations and other Monte Carlo codes~\cite{bib:mmc}, and found good agreement.

\section{Conclusions}
\label{sec:conclusions}
We presented the first search for magnetic monopoles ever performed with a UHECR detector, using the Pierre Auger Observatory. The particle showers produced by electromagnetic interactions of an ultrarelativistic monopole along its path through the atmosphere result in an energy deposit comparable to that of a UHECR, but with a very distinct profile which can be distinguished by the fluorescence detector. We have looked for such showers in the sample of hybrid events collected with Auger between 2004 and 2012, and no candidate was found. A 90\% C.L. upper limit on the flux of magnetic monopoles was placed, which is compared with results from previous experiments in Figure~\ref{fig:limit}.  Ours is the best limit for $\gamma \ge 10^9$, with a factor of ten improvement  for $\gamma \ge 10^{9.5}$. This result is valid for a broad class of intermediate-mass ultrarelativistic monopoles ($E_{\rm{mon}} \approx 10^{25}$~eV and  $ M \sim 10^{11} - 10^{16}$~eV/c$^2$) which may be present today as a relic of phase transitions in the early universe. Since the background - less than 0.1 event in the current data set - is not a limiting factor in the search, the upper bound will improve with the steadily increasing exposure of the Pierre Auger Observatory. 
\input{acknowledgments}

This work was supported in part by NSF grant PHY-1412261 and by the Kavli Institute for Cosmological Physics at the University of Chicago through grant NSF PHY-1125897 and an endowment from the Kavli Foundation and its founder Fred Kavli.
TF was supported by the Japan Society for the Promotion of Science Fellowship for Research Abroad H25-339.

\bibliography{monopole-search-prd}

\end{document}

%% file: revtex_authorlist.tex

\author{A.~Aab}
\affiliation{Universit\"at Siegen, Fachbereich 7 Physik -- Experimentelle Teilchenphysik, Germany}

\author{P.~Abreu}
\affiliation{Laborat\'orio de Instrumenta\c{c}\~ao e F\'\i{}sica Experimental de Part\'\i{}culas -- LIP and Instituto Superior T\'ecnico -- IST, Universidade de Lisboa -- UL, Portugal}

\author{M.~Aglietta}
\affiliation{Osservatorio Astrofisico di Torino (INAF), Torino, Italy}
\affiliation{INFN, Sezione di Torino, Italy}

\author{I.~Al Samarai}
\affiliation{Laboratoire de Physique Nucl\'eaire et de Hautes Energies (LPNHE), Universit\'es Paris 6 et Paris 7, CNRS-IN2P3, France}

\author{I.F.M.~Albuquerque}
\affiliation{Universidade de S\~ao Paulo, Inst.\ de F\'\i{}sica, S\~ao Paulo, Brazil}

\author{I.~Allekotte}
\affiliation{Centro At\'omico Bariloche and Instituto Balseiro (CNEA-UNCuyo-CONICET), Argentina}

\author{A.~Almela}
\affiliation{Instituto de Tecnolog\'\i{}as en Detecci\'on y Astropart\'\i{}culas (CNEA, CONICET, UNSAM), Centro At\'omico Constituyentes, Comisi\'on Nacional de Energ\'\i{}a At\'omica, Argentina}
\affiliation{Universidad Tecnol\'ogica Nacional -- Facultad Regional Buenos Aires, Argentina}

\author{J.~Alvarez Castillo}
\affiliation{Universidad Nacional Aut\'onoma de M\'exico, M\'exico}

\author{J.~Alvarez-Mu\~niz}
\affiliation{Universidad de Santiago de Compostela, Spain}

\author{M.~Ambrosio}
\affiliation{INFN, Sezione di Napoli, Italy}

% ...

\collaboration{The Pierre Auger Collaboration}
\email{auger_spokespersons@fnal.gov}
\homepage{http://www.auger.org}
\noaffiliation

%% file: acknowledgments.tex

\section*{Acknowledgments}

\begin{sloppypar}
The successful installation, commissioning, and operation of the Pierre Auger Observatory would not have been possible without the strong commitment and effort from the technical and administrative staff in Malarg\"ue. We are very grateful to the following agencies and organizations for financial support:
\end{sloppypar}

\begin{sloppypar}
Comisi\'on Nacional de Energ\'\i{}a At\'omica, Agencia Nacional de Promoci\'on Cient\'\i{}fica y Tecnol\'ogica (ANPCyT), Consejo Nacional de Investigaciones Cient\'\i{}ficas y T\'ecnicas (CONICET), Gobierno de la Provincia de Mendoza, Municipalidad de Malarg\"ue, NDM Holdings and Valle Las Le\~nas, in gratitude for their continuing cooperation over land access, Argentina; the Australian Research Council; Conselho Nacional de Desenvolvimento Cient\'\i{}fico e Tecnol\'ogico (CNPq), Financiadora de Estudos e Projetos (FINEP), Funda\c{c}\~ao de Amparo \`a Pesquisa do Estado de Rio de Janeiro (FAPERJ), S\~ao Paulo Research Foundation (FAPESP) Grants No.\ 2010/07359-6 and No.\ 1999/05404-3, Minist\'erio de Ci\^encia e Tecnologia (MCT), Brazil; Grant No.\ MSMT CR LG15014, LO1305 and LM2015038 and the Czech Science Foundation Grant No.\ 14-17501S, Czech Republic; Centre de Calcul IN2P3/CNRS, Centre National de la Recherche Scientifique (CNRS), Conseil R\'egional Ile-de-France, D\'epartement Physique Nucl\'eaire et Corpusculaire (PNC-IN2P3/CNRS), D\'epartement Sciences de l'Univers (SDU-INSU/CNRS), Institut Lagrange de Paris (ILP) Grant No.\ LABEX ANR-10-LABX-63, within the Investissements d'Avenir Programme Grant No.\ ANR-11-IDEX-0004-02, France; Bundesministerium f\"ur Bildung und Forschung (BMBF), Deutsche Forschungsgemeinschaft (DFG), Finanzministerium Baden-W\"urttemberg, Helmholtz Alliance for Astroparticle Physics (HAP), Helmholtz-Gemeinschaft Deutscher Forschungszentren (HGF), Ministerium f\"ur Wissenschaft und Forschung, Nordrhein Westfalen, Ministerium f\"ur Wissenschaft, Forschung und Kunst, Baden-W\"urttemberg, Germany; Istituto Nazionale di Fisica Nucleare (INFN),Istituto Nazionale di Astrofisica (INAF), Ministero dell'Istruzione, dell'Universit\'a e della Ricerca (MIUR), Gran Sasso Center for Astroparticle Physics (CFA), CETEMPS Center of Excellence, Ministero degli Affari Esteri (MAE), Italy; Consejo Nacional de Ciencia y Tecnolog\'\i{}a (CONACYT) No.\ 167733, Mexico; Universidad Nacional Aut\'onoma de M\'exico (UNAM), PAPIIT DGAPA-UNAM, Mexico; Ministerie van Onderwijs, Cultuur en Wetenschap, Nederlandse Organisatie voor Wetenschappelijk Onderzoek (NWO), Stichting voor Fundamenteel Onderzoek der Materie (FOM), Netherlands; National Centre for Research and Development, Grants No.\ ERA-NET-ASPERA/01/11 and No.\ ERA-NET-ASPERA/02/11, National Science Centre, Grants No.\ 2013/08/M/ST9/00322, No.\ 2013/08/M/ST9/00728 and No.\ HARMONIA 5 -- 2013/10/M/ST9/00062, Poland; Portuguese national funds and FEDER funds within Programa Operacional Factores de Competitividade through Funda\c{c}\~ao para a Ci\^encia e a Tecnologia (COMPETE), Portugal; Romanian Authority for Scientific Research ANCS, CNDI-UEFISCDI partnership projects Grants No.\ 20/2012 and No.194/2012 and PN 16 42 01 02; Slovenian Research Agency, Slovenia; Comunidad de Madrid, Fondo Europeo de Desarrollo Regional (FEDER) funds, Ministerio de Econom\'\i{}a y Competitividad, Xunta de Galicia, European Community 7th Framework Program, Grant No.\ FP7-PEOPLE-2012-IEF-328826, Spain; Science and Technology Facilities Council, United Kingdom; Department of Energy, Contracts No.\ DE-AC02-07CH11359, No.\ DE-FR02-04ER41300, No.\ DE-FG02-99ER41107 and No.\ DE-SC0011689, National Science Foundation, Grant No.\ 0450696, The Grainger Foundation, USA; NAFOSTED, Vietnam; Marie Curie-IRSES/EPLANET, European Particle Physics Latin American Network, European Union 7th Framework Program, Grant No.\ PIRSES-2009-GA-246806; and UNESCO.
\end{sloppypar}